\documentclass[conference]{IEEEtran}
\usepackage{graphicx}
\usepackage{amsmath, amssymb}
\usepackage{cite}
\usepackage{hyperref}
\usepackage{caption}

\title{Diffusion Models on the Edge: Challenges, Optimizations, and Applications}

\author{\IEEEauthorblockN{Dongqi Zheng}
\IEEEauthorblockA{\textit{Apple Inc.} \\}
}

\begin{document}

\maketitle

\begin{abstract}
Diffusion models have shown remarkable capabilities in generating high-fidelity data across modalities such as images, audio, and video. However, their computational intensity makes deployment on edge devices a significant challenge. This survey explores the foundational concepts of diffusion models, identifies key constraints of edge platforms, and synthesizes recent advancements in model compression, sampling efficiency, and hardware-software co-design to make diffusion models viable on edge devices. We also review promising applications and suggest future research directions.
\end{abstract}


\section*{Impact Statement}
Diffusion models have shown impressive results in generating high-quality images, audio, and video, but their high computational cost has restricted their use to powerful cloud servers. This paper bridges that gap by enabling such models to run on local, low-power devices like smartphones, IoT sensors, and wearables. This advancement could significantly reduce energy consumption and data privacy concerns associated with cloud computing, which is vital from both an environmental and social standpoint. Technologically, it opens the door for real-time, personalized AI experiences at the edge—without internet reliance—benefiting applications in healthcare, education, assistive devices, and industrial monitoring. Economically, it could lower operational costs for businesses and expand AI accessibility in underserved regions. The paper thus lays the groundwork for the next generation of efficient, private, and sustainable edge intelligence.

\section{Introduction}
Diffusion models have rapidly become a cornerstone in the field of generative modeling, outperforming traditional methods in high-dimensional data synthesis such as image generation, audio synthesis, and video prediction. These models work by reversing a progressive noising process, enabling them to generate highly realistic outputs. As the demand for intelligent and personalized applications increases, there is a growing interest in bringing such powerful generative capabilities to the edge, where computation is performed locally on resource-constrained devices like smartphones, IoT sensors, and wearables.

Deploying diffusion models on the edge presents significant advantages, including enhanced privacy (as data stays on the device), reduced latency (by avoiding cloud round-trips), and increased personalization. However, realizing these benefits is non-trivial due to the massive computational and memory requirements of traditional diffusion models. This survey provides a comprehensive overview of the diffusion model landscape with a focus on their adaptation to edge environments. We discuss their underlying principles, identify challenges in their deployment, review optimization strategies and co-design techniques, examine practical edge applications, and conclude with potential future directions in this domain.

\section{Background on Diffusion Models}
Diffusion models are a class of generative models that learn to generate data by reversing a diffusion process that progressively corrupts data with noise. The most widely adopted variant, Denoising Diffusion Probabilistic Models (DDPMs), was introduced by \cite{ho2020ddpm}. These models consist of two main processes: the forward process, where Gaussian noise is added to data over a fixed number of steps, and the reverse process, where a neural network is trained to iteratively remove this noise and recover the original data.

The reverse process is modeled by a U-Net, a deep convolutional neural network designed to predict the noise component added in each step. The learning objective typically minimizes the mean squared error between the predicted and actual noise. Despite the simplicity of the training objective, diffusion models have demonstrated impressive results, surpassing GANs in image quality and sample diversity.

\begin{figure}[h]
  \centering
  \includegraphics[width=0.48\textwidth]{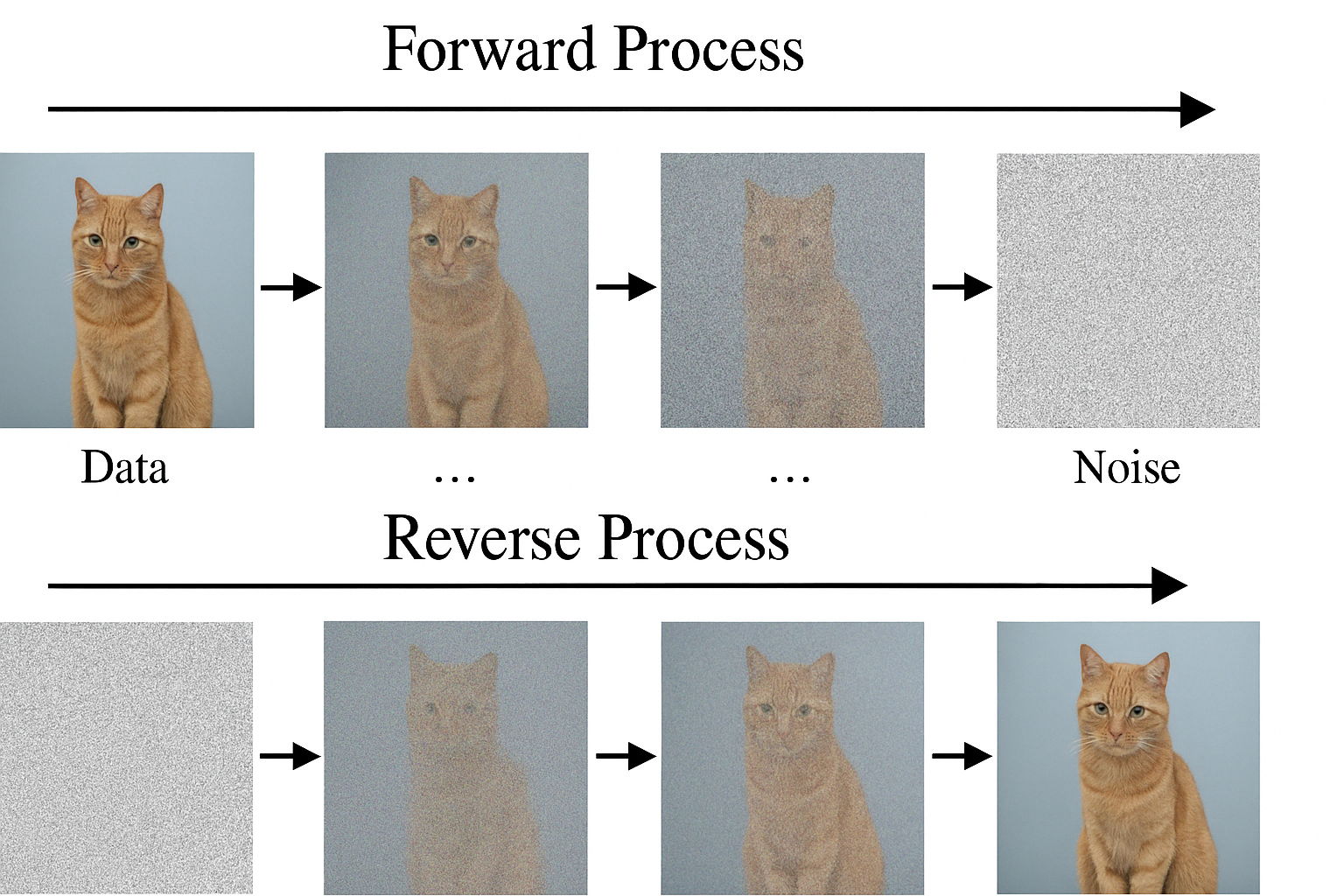}
  \caption{Illustration of the forward and reverse diffusion process used in generative models.}
  \label{fig:diffusion_cat}
\end{figure}

Variants of diffusion models have emerged to address specific needs. 
\textbf{Latent Diffusion Models (LDMs)} apply the diffusion process in a compressed latent space, significantly reducing computational costs~\cite{rombach2022ldm}. 
\textbf{Conditional diffusion models} allow generation to be guided by class labels, text prompts, or other modalities~\cite{dhariwal2021diffusion}. 
Additionally, architectures like \textbf{ControlNet} enable precise control over outputs using structured input such as edge maps or poses~\cite{zhang2023controlnet}. 
These advancements have made diffusion models versatile and capable, but they still remain computationally expensive and unsuitable for direct deployment on low-power devices.

\begin{table*}[t]
\small
\centering
\caption{Comparison of Edge Platforms for Diffusion Model Deployment}
\label{tab:edge_platforms}
\begin{tabular}{|l|c|c|c|c|}
\hline
\textbf{Metric} & \textbf{MCU} & \textbf{SoC} & \textbf{NPU} & \textbf{FPGA} \\
\hline
Memory & 64KB–1MB & 1–12GB & 512MB–4GB & 512MB–4GB \\
\hline
Clock Speed & 48–480MHz & 1–3GHz & 0.5–2GHz & Configurable \\
\hline
Power Consumption & 10–100mW & 1–3W & 0.5–2W & 1–5W \\
\hline
Compute Throughput & Low & High & Very High & Medium–High \\
\hline
Flexibility & Low & Medium & Low–Medium & Very High \\
\hline
Ease of Deployment & High & High & Medium & Low \\
\hline
Real-time Capability & Excellent & Good & Good & Good \\
\hline
Example Chips & STM32, GAP9 & A16, Snapdragon & EdgeTPU, K210 & Zynq, Intel MAX \\
\hline
\end{tabular}
\end{table*}

\section{Overview of Edge Computing Platforms}

\subsection{Edge Computing Paradigm}
Edge computing shifts data processing away from centralized servers to local devices closer to the data source. This architectural shift aims to reduce latency, enhance privacy, and lower bandwidth usage. Edge devices execute AI tasks directly on-device, providing fast responses and operating even without internet access. These qualities are essential for real-time applications like autonomous navigation, smart cameras, and wearable health monitors.

\subsection{Types of Edge Platforms}
Edge platforms are diverse and include:

\textbf{Microcontrollers (MCUs)}: Devices like STM32 or Cortex-M families offer extremely limited memory (typically in the kilobyte to megabyte range) and operate at low clock speeds (tens to hundreds of MHz). These are widely used in deeply embedded systems where energy efficiency is critical.

\textbf{Mobile SoCs}: Integrated platforms such as Apple’s A-series or Qualcomm’s Snapdragon feature heterogeneous computing elements including CPUs, GPUs, and NPUs. These offer significantly more compute power than MCUs while maintaining energy efficiency and are optimized for battery-powered devices like smartphones and tablets.

\textbf{Dedicated NPUs}: Neural Processing Units like Google’s EdgeTPU or Kendryte K210 are specialized for deep learning tasks. These accelerators handle matrix operations efficiently and are being increasingly integrated into edge AI devices.

\textbf{FPGAs}: Field-programmable gate arrays offer reconfigurable logic, enabling tailored acceleration for specific models. While offering great flexibility, they typically require more engineering effort and may be less power-efficient than NPUs.

\subsection{System-Level Constraints}
Each type of edge platform imposes distinct constraints:

- \textbf{Compute Throughput}: Often limited to a few TOPS (tera operations per second) or less, requiring models to be highly optimized.

- \textbf{Memory Bandwidth and Capacity}: Typically constrained to a few megabytes of SRAM or shared DRAM, with strict latency limits.

- \textbf{Power Budget}: Devices often operate under a sub-watt power envelope, making traditional high-performance models infeasible without substantial optimization.

- \textbf{Thermal Management}: Passive cooling requirements mean that sustained high-throughput workloads must be avoided or throttled.

\subsection{Deployment Frameworks and Software Tooling}
To bridge the gap between high-performance diffusion models and these constrained devices, various software frameworks have emerged:

- \textbf{TensorFlow Lite Micro} is optimized for microcontrollers and supports integer quantization and memory-efficient kernels.

- \textbf{ONNX Runtime} offers cross-platform model execution and supports accelerators through custom backends.

- \textbf{CoreML} is designed for Apple devices and integrates well with mobile SoC hardware accelerators.

- \textbf{Apache TVM} provides compilation and optimization pipelines targeting NPUs and FPGAs with customizable operator kernels.

These frameworks are crucial in mapping neural networks to resource-constrained edge environments. They handle quantization, memory layout adjustments, and low-level operator fusion. However, most diffusion models in their vanilla form are still too large or slow for direct deployment, especially when hundreds of sequential steps are required. \cite{warden2019tinyml, mlperftiny, chen2020edgeai}

\section{Challenges in Deploying Diffusion Models on Edge Devices}
The deployment of diffusion models on edge devices faces a set of multifaceted challenges. First and foremost is the computational complexity of the reverse diffusion process. Traditional diffusion models require hundreds to thousands of denoising steps to generate a single sample. Each step involves a full forward pass through a deep neural network, making the entire process computationally intensive and time-consuming.

Secondly, diffusion models consume a significant amount of memory. The U-Net architecture used in most implementations requires large intermediate feature maps to be stored during each step of the reverse process. This memory requirement often exceeds the capacity of microcontrollers and embedded platforms, necessitating creative memory management or model simplification.

Another major challenge is latency. Real-time applications, such as video generation or speech synthesis, require rapid response times that are incompatible with the long inference time of standard diffusion models. Reducing latency without significantly degrading output quality remains a major research goal.

Power consumption is also a limiting factor. Edge devices are often battery-powered and operate in environments where thermal dissipation is limited. The iterative and computationally expensive nature of diffusion models leads to high power draw, which is unsuitable for such devices.

Lastly, diffusion models are not inherently optimized for the characteristics of edge hardware. Unlike single-shot models such as CNNs, diffusion models require recurrent computation and often use operations that are inefficient on edge accelerators. This mismatch necessitates rethinking both the model architecture and the deployment strategy to suit the hardware.

\section{Model Optimization Techniques for Edge Deployment}

\subsection{Sampling Acceleration}
One of the primary bottlenecks in diffusion models is the number of denoising steps required to generate high-quality outputs. Traditional DDPM models require up to 1000 steps. To address this, acceleration techniques like DDIM (Denoising Diffusion Implicit Models) and DPM-Solver have been developed \cite{nichol2021improved,lu2022dpm}. DDIM provides a non-Markovian deterministic sampling approach that can achieve generation in as few as 20–50 steps without retraining. DPM-Solver, on the other hand, leverages numerical solvers for ordinary differential equations (ODEs) to approximate the reverse diffusion process. These methods drastically reduce latency and energy consumption, making real-time deployment on the edge more feasible.

\subsection{Architectural Simplification}
Reducing the architectural complexity of diffusion models is essential for fitting within the compute and memory constraints of edge devices. Mobile-friendly U-Net variants such as MobileU-Net and Tiny-Diffusion use efficient operations like depthwise separable convolutions and inverted residual blocks, inspired by MobileNet and EfficientNet \cite{howard2019searching,li2023mobilediffusion}. These structures maintain representational capacity while drastically cutting the number of parameters and FLOPs. Researchers also explore trimming channel widths and reducing the number of layers without severely compromising output quality. Techniques like neural architecture search (NAS) have also been applied to discover minimal diffusion backbones suitable for MCUs and NPUs.

\subsection{Latent Space Diffusion}
Latent Diffusion Models (LDMs) achieve significant efficiency by operating in a compressed feature space, rather than the full-resolution pixel space \cite{rombach2022ldm}. These models use a pre-trained autoencoder (typically a variational autoencoder or VQGAN) to encode inputs into a latent space, where the denoising process takes place. This approach reduces both spatial and channel dimensions, enabling faster and more memory-efficient inference. Once generation is complete, the output is decoded back into the image domain. Since the diffusion network operates on smaller tensors, it is significantly more suitable for edge environments where SRAM and memory bandwidth are limited.

\subsection{Quantization and Pruning}
Model compression is crucial for edge deployment, and quantization is one of the most effective techniques. Post-training quantization reduces the bit-width of weights and activations, typically to 8-bit integers (INT8) or even lower, while preserving the statistical distribution of the data \cite{jacob2018quantization}. Quantization-aware training (QAT) can further minimize accuracy loss by simulating quantization effects during training. Pruning complements quantization by removing redundant or low-importance weights and filters, thereby reducing the overall model size and computational load. Techniques such as magnitude pruning, structured pruning, and lottery ticket hypothesis are actively used in diffusion models to remove inefficiencies.

\subsection{Knowledge Distillation}
Knowledge distillation offers another path for model size reduction by training a compact “student” model to replicate the behavior of a large “teacher” model \cite{hinton2015distilling}. In the context of diffusion models, the teacher could be a large U-Net trained over hundreds of steps, while the student learns to mimic its predictions over fewer steps or with a smaller architecture. This technique not only reduces inference time and memory usage but can also help transfer generalization from the teacher to the student. Distilled diffusion models have been shown to match or exceed the quality of their larger counterparts when trained carefully.

\subsection{Operator Fusion and Kernel Optimization}
Edge platforms benefit immensely from operation-level optimizations. Operator fusion combines multiple consecutive layers—such as convolution, batch normalization, and ReLU—into a single computational kernel. This reduces memory read/write operations, improves data locality, and leverages vectorized or parallel instructions \cite{chen2018tvm}. Many compilers, such as TVM, Glow, and proprietary SDKs, support automatic fusion and operator rewriting. Kernel optimization also includes tuning loop structures, adjusting memory access patterns, and exploiting hardware features like SIMD, tensor cores, or DMA for zero-copy transfers. For NPUs and DSPs, hand-optimized or auto-generated kernels (via MLIR or AutoTVM) can bring 2–5x speedups compared to naive deployment.

Together, these model optimization techniques form a toolkit that allows developers to fit complex generative pipelines like diffusion models within the tight compute, memory, and power envelopes of edge AI platforms.

\section{Hardware-Software Co-Design for Efficient Inference}

\subsection{Co-Design Fundamentals}
The essence of hardware-software co-design is to jointly optimize the model architecture and deployment stack to maximize performance under hardware constraints such as compute capacity, memory bandwidth, and power budget. Unlike traditional development flows where software is developed independently from hardware, co-design ensures that models are adapted to the quirks of the target hardware platform—be it a microcontroller, an NPU, or an FPGA. This process includes quantization-aware training, memory-aware architecture search, tiling strategies, and instruction-level fusion \cite{lin2020mcunet, cai2020once}.

\subsection{Scheduling and Memory Optimization}
Efficient scheduling is essential for operating within the limited on-chip SRAM of edge platforms. A diffusion model involves iterative computation across many steps, each requiring a forward pass through a neural network. These steps can be optimized by designing a schedule that reuses intermediate memory buffers, overlaps computation with data movement (e.g., via DMA), and minimizes memory fragmentation \cite{chen2018tvm}.

To avoid SRAM overflow, developers commonly use tiling strategies where large feature maps are broken into smaller blocks that can be loaded, processed, and offloaded incrementally. Coupling this with memory pooling and static memory allocation techniques ensures predictable and efficient memory use. SRAM reuse is often necessary, especially on MCUs like STM32H7 which provide only hundreds of KB of RAM.

\subsection{Quantization-Aware Layouts and Operator Fusion}
Quantization-aware layouts reorganize tensors to better match the access patterns of the underlying accelerator. For example, some NPUs prefer channel-last formats to maximize vectorization, while others optimize for memory-aligned access. Understanding this allows developers to pre-transform the layout and pack tensors into native formats such as NCHW4 or int8-encoded tiles \cite{armethosu55}.

Operator fusion is another major lever. In diffusion models, many operations are composable: convolution, batch normalization, activation, and dropout can be fused into a single kernel to reduce latency and memory access overhead. Compilers like TVM, Glow, and proprietary SDKs (e.g., NPU SDKs from Kendryte or ARM Ethos-U) support auto-fusion and scheduling transformations to exploit these fusions \cite{chen2018tvm}.

\subsection{Runtime Execution Frameworks and Compilers}
Deployment frameworks such as \textbf{CMSIS-NN}, \textbf{TensorFlow Lite Micro}, \textbf{TVM}, and \textbf{ONNX Runtime} are often used to compile and optimize models for embedded deployment. TVM, for instance, performs auto-tuning and kernel-level scheduling, while CMSIS-NN is manually optimized with SIMD intrinsics for ARM Cortex-M CPUs \cite{gadi2021cmsisnn}.

Compiler passes can include static quantization insertion (transforming float models into low-bit integer ones), memory buffer coalescing (reusing memory blocks for different stages), instruction pipelining (overlapping compute with load/store), and inference graph simplification (folding constants, eliminating redundant operations).

\subsection{Profiling and Performance Tuning}
To analyze and fine-tune inference behavior, developers employ low-level profiling tools:

\textbf{Arm Streamline} provides real-time tracing of CPU, GPU, and NPU workloads, visualizing performance bottlenecks such as cache misses and core underutilization.

\textbf{GAPBench} is tailored for GAP processors, enabling profiling of RISC-V-based edge computing workloads, memory usage, and inter-core DMA transfers.

\textbf{Vendor SDK Profilers} like Google's EdgeTPU profiler or Kendryte K210 toolchain help pinpoint operator-level latency and estimate power draw, facilitating device-aware tuning \cite{kendryte2021}.

These tools provide cycle-level and energy-level breakdowns, helping developers detect hotspots, underutilized compute units, and bandwidth bottlenecks.

\subsection{Case Study: Platform-Specific Optimization}
On the STM32H7 (ARM Cortex-M7, no NPU), developers often use CMSIS-NN to hand-optimize convolution layers with SIMD instructions, using 8-bit quantized weights to fit within SRAM. Layers are tiled spatially and offloaded to SRAM in a double-buffered pattern.

On the GAP9 platform, RISC-V cores are paired with a cluster-based architecture for parallelism. Developers use GAP AutoTiler to partition computation and manage DMA explicitly. When paired with a Kendryte K210-like NPU, inference graphs must be transformed into fixed operator sequences supported by the hardware scheduler.

For ARM Ethos-U55, TensorFlow Lite for Microcontrollers (TFLM) integrates directly with the NPU backend. Developers quantize the model (via TFLite Converter), use the Vela compiler to map subgraphs to Ethos-U, and the runtime handles execution split between CPU and NPU \cite{armethosu55}.

\subsection{Challenges and Future Trends}
Despite these advances, co-design still faces multiple practical and research-level challenges:

\textbf{Lack of hardware abstraction}: Many NPUs lack a unified execution abstraction, forcing developers to write custom kernels or graph compilers per device.

\textbf{Limited open-source tools}: Performance tuning is often restricted to proprietary SDKs, hindering transparency and portability.

\textbf{Latency variability}: Unpredictable memory access patterns can cause latency spikes, particularly when caching and bandwidth are not explicitly controlled.

\textbf{Poor support for iterative models}: Most compilers and frameworks are optimized for feedforward networks and struggle to optimize the recurrent or iterative nature of diffusion models.

Emerging solutions include differentiable co-design frameworks that combine hardware-aware NAS with model training, instruction set customization for efficient iterative computation, and joint compiler-hardware stack co-optimization \cite{lin2020mcunet, cai2020once, chen2018tvm}.

\section{Applications of Diffusion Models on the Edge}
Despite their resource demands, diffusion models have promising use cases in edge environments. In mobile photography and augmented reality, they can be used for tasks such as super-resolution, denoising, and inpainting. For instance, a diffusion model running on a smartphone could enhance a low-light photo or generate missing portions of an image in real time.

In audio applications, diffusion models can generate high-quality speech or sound effects on wearable devices. Trained on datasets like VCTK or LibriTTS, these models can produce personalized voices or ambient soundscapes for headphones, hearing aids, or sleep devices.

Augmented reality applications benefit from diffusion models through background augmentation, virtual object placement, and scene generation. These tasks require rapid and localized image synthesis, which can be achieved with optimized diffusion models.

In healthcare and IoT, diffusion models can generate or reconstruct physiological signals for anomaly detection or signal enhancement. For example, a wearable health monitor might use a diffusion model to denoise ECG signals or generate missing data during transmission interruptions.

Finally, creative applications such as sketch-to-image conversion, on-device design tools, and personalized filters can benefit from lightweight generative models. These allow artists and designers to experiment with AI-powered creation without relying on cloud connectivity.

\section{Benchmarking and Evaluation Metrics}

\subsection{Fidelity Metrics for Generated Outputs}
Evaluating the output quality of diffusion models is crucial for generative tasks. The \textbf{Fréchet Inception Distance (FID)} is a standard metric that compares the statistics of generated and real images in the feature space of a pretrained Inception network. Lower FID values indicate more realistic outputs. For image restoration and super-resolution tasks, \textbf{Peak Signal-to-Noise Ratio (PSNR)} and \textbf{Structural Similarity Index (SSIM)} are widely used. PSNR measures the pixel-level difference between generated and ground-truth images, while SSIM accounts for luminance, contrast, and structure similarity, better reflecting human perception.

\subsection{Latency and Throughput}
\textbf{Latency} is measured as the time required to generate one complete sample (e.g., one image, one audio segment). For real-time applications, especially on mobile and embedded devices, maintaining low latency (e.g., $<50$ ms/sample) is essential. \textbf{Throughput} refers to the number of samples that can be generated per second and is especially relevant in batch inference use cases, such as AR rendering or background generation. Benchmarking latency involves measuring both average and worst-case runtimes under realistic deployment scenarios.

\subsection{Energy and Power Consumption}
Power efficiency is another critical dimension in edge inference. \textbf{Energy per sample} (in Joules) captures the total energy drawn to generate a single output, while \textbf{average power draw} (in Watts) is measured over the inference period. These metrics are particularly relevant for battery-powered or energy-harvesting edge devices. Energy profiling tools such as \textit{Arm Streamline}, \textit{GAPBench}, and device-specific SDKs enable developers to capture these metrics with granularity \cite{banbury2021benchmarking, lin2020mcunet}.

\subsection{Memory Usage and Model Size}
The peak \textbf{RAM usage} during inference reflects the model’s memory footprint and determines if it can run within the on-chip SRAM of target devices. SRAM on many MCUs is limited to hundreds of KB to a few MB. Additionally, \textbf{model size after compression} (e.g., quantization, pruning) determines storage requirements and influences deployment strategy (e.g., in-flash vs. streamed). Tools such as 	extit{TensorFlow Lite Converter} and 	extit{TVM} provide model size estimation after compilation.

\subsection{Standard Benchmarks and Tools}
To promote reproducibility and cross-platform comparison, benchmarking suites like \textbf{MLPerf Tiny} offer standardized test scenarios for MCUs and constrained devices \cite{mlperftiny}. These benchmarks report latency, energy, and accuracy across several tasks (e.g., keyword spotting, visual wake words). Meanwhile, profiling tools like \textit{Arm Streamline}, \textit{EdgeML Dashboard}, and \textit{GAPBench} offer visual and programmable ways to trace model behavior in terms of memory usage, kernel time, and power draw.

These metrics provide a multidimensional view of model performance, helping developers balance trade-offs in fidelity, efficiency, and deployability when adapting diffusion models to constrained environments.

\section{Future Directions}
Several promising directions can extend the reach of diffusion models on the edge. Cross-modal diffusion models, which convert between text, image, and audio domains, are becoming increasingly popular. Efficient edge-friendly implementations of text-to-image or text-to-audio diffusion could unlock powerful real-time assistant tools and creative applications.

Federated learning and privacy-preserving techniques allow diffusion models to be trained or fine-tuned locally, enabling on-device personalization without compromising user data. These techniques are critical for applications in healthcare, security, and personal media generation.

\begin{table*}[h]
\caption{Key Evaluation Metrics for Edge-Deployed Diffusion Models}
\centering
\renewcommand{\arraystretch}{1.2}
\begin{tabular}{|p{2.6cm}|p{2.3cm}|p{2.8cm}|}
\hline
\textbf{Category} & \textbf{Metric} & \textbf{Purpose} \\
\hline
Output Quality & FID, PSNR, SSIM & Measures visual or perceptual fidelity of generated outputs \\
\hline
Latency & ms/sample & Indicates responsiveness in real-time applications \\
\hline
Throughput & samples/sec & Assesses performance under high-volume inference \\
\hline
Energy Efficiency & Joules/sample & Quantifies energy cost per generated sample \\
\hline
Power Consumption & Watts (avg/peak) & Evaluates thermal and power requirements \\
\hline
Memory Usage & Peak RAM (MB) & Checks fit within SRAM or memory limits \\
\hline
Model Size & MB & Determines on-device storage feasibility after compression \\
\hline
Benchmark Suite & MLPerf Tiny or EdgeML & Provides standardized cross-platform evaluation \\
\hline
\end{tabular}
\label{tab:metrics}
\end{table*}

Event-driven inference and sparse sampling strategies can be explored to reduce redundant computation. Instead of executing the full model at regular intervals, inference can be triggered by changes in input or environment, improving efficiency.

AutoML and architecture search methods tailored for edge environments can be used to discover optimal model configurations for specific hardware platforms. These tools automate the process of designing diffusion models that balance accuracy and efficiency.

Finally, developing diffusion models with hardware in mind from the outset—co-optimizing for instruction sets, parallelism, and memory hierarchy—can result in truly edge-native generative models that meet the demanding requirements of real-world applications.

\section{Conclusion}
Diffusion models have revolutionized generative AI, achieving remarkable results in domains ranging from image synthesis to speech generation. While their computational requirements have traditionally limited their deployment to server-class hardware, recent advances in model optimization, sampling efficiency, and hardware-software co-design are bringing diffusion models closer to practical use on edge devices.

This survey has explored the foundations of diffusion models, the challenges of edge deployment, the state-of-the-art optimization techniques, and the breadth of potential applications. We have also identified promising future directions that could further improve the feasibility of running diffusion models on constrained platforms.

\bibliographystyle{IEEEtran}
\bibliography{diffusion_edge_survey_updated}

\end{document}